\documentclass{article}

\usepackage{arxiv}

\usepackage[utf8]{inputenc} 
\usepackage[T1]{fontenc}    
\usepackage{url}            
\usepackage{booktabs}       
\usepackage{amsfonts}       
\usepackage{nicefrac}       
\usepackage{microtype}      
\usepackage{graphicx}       
\title{A 0.16pJ/bit Recurrent Neural Network Based PUF for Enhanced Machine Learning Attack Resistance}

\author{ Nimesh Shah\\
Nanyang Technological University\\
Singapore 639798\\
\texttt{shah0042@e.ntu.edu.sg}
\\\And
 Manaar Alam\\
Indian Institute of Technology\\
Kharagpur\\
\texttt{alam.manaar@iitkgp.ac.in}\And
Durga Prasad Sahoo\\
Robert Bosch Engineering\\
India\\
\texttt{dpsahoo.cs@gmail.com}\And
Debdeep Mukhopadhyay\\
Indian Institute of Technology\\
Kharagpur\\
\texttt{debdeep@cse.iitkgp.ac.in} \\
\And
Arindam Basu\\
Nanyang Technological University\\
Singapore 639798\\
\texttt{arindam.basu@ntu.edu.sg} \\
}

\begin{document}
\maketitle

\begin{abstract}
Physically Unclonable Function (PUF) circuits are finding widespread use due to increasing adoption of IoT devices. However, the existing strong PUFs such as Arbiter PUFs (APUF) and its compositions are susceptible to machine learning (ML) attacks because the challenge-response pairs have a linear relationship. In this paper, we present a Recurrent-Neural-Network PUF (RNN-PUF) which uses a combination of feedback and XOR function to significantly improve resistance to ML attack, without significant reduction in the reliability. ML attack is also partly reduced by using a shared comparator with offset-cancellation to remove bias and save power. From simulation results, we obtain ML attack accuracy of 62\% for different ML algorithms, while reliability stays above 93\%. This represents a 33.5\% improvement in our Figure-of-Merit. Power consumption is estimated to be $12.3\mu$W with energy/bit of $\approx 0.16$pJ.
\end{abstract}

\section{Introduction}
Device connectivity in the Internet of Things (IoT) is predicted to increase exponentially while electronics usage in automobiles has also increased significantly. With the arrival of self-driving-automobile technology, the role of hardware security will only get more important. PUFs provide a safe way to generate unique identification(ID) or authentication bits, compared with non-volatile memories that can be hacked using widely known techniques. Both analog and digital circuit techniques have been used to design PUFs \cite{alioto:isscc,stanzione:jssc,ma:aspdac}, with a preference for sub-threshold analog techniques for low power operation. There are two kinds of PUFs, weak and strong \cite{maes}, with the weak one used solely for ID, while the strong one being used for authentication and/or ID. Strong PUFs have a large challenge-response pair (CRP) space such that an attacker cannot predict the response of some arbitrary challenge even though he has knowledge of many other CRPs. However, several mathematical models have been developed for existing strong PUFs, largely using machine learning (ML) techniques to learn the expected patterns of the PUF's CRPs\cite{ruhrmair:acm}. Alternative design strategies using analog components have been proposed like in \cite{supreet:vlsi,orshansky:vlsi} with improved resistance than the classic Arbiter PUF. In this paper, we present a technique to significantly decrease the machine learning attack accuracy (MLAA) more specifically for analog PUFs using neural-networks (NN), without being heavily penalized on reliability. Our RNN-PUF uses a combination of recurrence, or feedback, and XOR to generate a response whose relationship with the particular challenge is obscured to the attacker due to the randomness provided by the previous response.

Our contributions in this work are as follows:
\begin{itemize}
\item We first show high MLAA of a recently proposed analog NN-PUF thus showing that analog PUFs are not inherently robust against ML attacks.
\item We propose a simple method to reduce MLAA for the NN-PUF at the cost of more area.
\item We propose the RNN-PUF that has both low MLAA and low area.
\item We propose a design guide for the RNN-PUF where area and MLAA/reliability are traded-off.
\end{itemize}

The paper is organized as follows: Section \ref{sec:nn-puf} describes the conventional PUF based on current mirror arrays (CMA). Section \ref{sec:rnnpuf} introduces the RNN-PUF, its operational details and its reliability/accuracy. Section \ref{sec:guide} discusses the design guide, while Section \ref{sec:pwr} presents power consumption simulations and comparison with state-of-art. Finally, Section \ref{sec:conc} concludes the paper.

\section{The Conventional NN-PUF}
\label{sec:nn-puf}
\begin{figure}
\centering
\includegraphics[clip=true,scale=0.35]{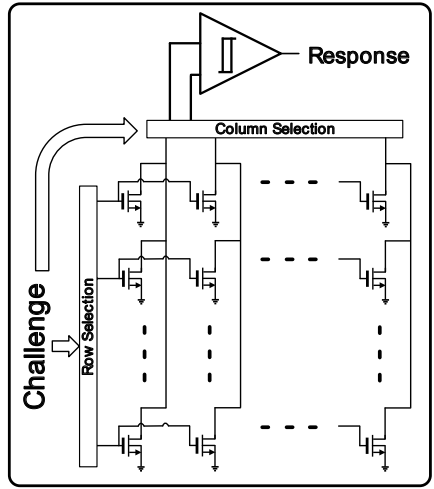}
\caption{Architecture of Conventional NN-PUF}
\label{convpuf}
\end{figure}

The conventional PUF without recurrence is based on a $128\times128$ Neural-Network PUF \cite{wang:tcas} using CMA with high mismatch. We choose the NN-PUF as the base PUF because it can be reused for ML, another important functionality becoming ubiquitous in IoT \cite{whatmough:isscc}. Also, it has the least area per bit reported so far. Similar arrays may be constructed out of the current mirrors reported in \cite{orshansky:vlsi,rkumar:host}. Figure \ref{convpuf} shows the schematic of the conventional NN-PUF. A 1-bit response is generated by comparing the output current in a column pair specified by the column challenge. The row challenge bits selects which row of transistors are active. The current values for each column of the column pair are summed up, and the two results are sent to the comparator to generate $1$($I^C_{col,A}>I^C_{col,B}$) or $0$($I^C_{col,A}<I^C_{col,B}$) where $I^C_{col,A}$ denotes the current output for column $A$ given a challenge $C$. The only difference is we show a shared comparator while \cite{wang:tcas} used current controlled oscillators (CCO) to digitize the current values. We note that the maximum number of possible column pairs is ${n_{col}\choose 2}$ where $n_{col}$ is the number of columns. However we can only get $\frac{n_{col}}{2}$ independent column pairs, because if we know that $I^C_{col,X}>I^C_{col,Y}$ and $I^C_{col,Y}>I^C_{col,Z}$, then we know for sure that $I^C_{col,X}>I^C_{col,Z}$ \cite{maes}. 

\subsection{Conventional NN-PUF Reliability}
In order to improve reliability of the NN-PUF, it was shown in \cite{wang:tcas} that we can use a dynamic threshold control, $V_{ctrl}$, but it comes at the cost of discarding more CRPs. In that scheme, a response is considered valid if and only if $\vert I^C_{col,X} - I^C_{col,Y}\vert > V_{ctrl}$. If invalid, a different challenge has to be used. 

\begin{figure}
\centering
\includegraphics[clip=true,scale=0.35]{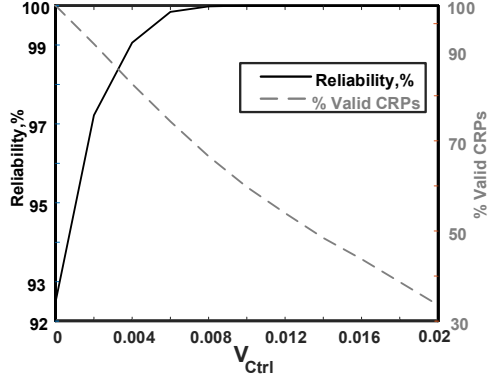}
\caption{Conventional NN-PUF Reliability variation against $V_{ctrl}$}
\label{convrel}
\end{figure}

Figure \ref{convrel} shows the reliability of the conventional PUF against $V_{ctrl}$, for $5000$ CRPs. Reliability is defined as the probability of 1-bit output being correct across different operating conditions such as temperature/voltage/time etc. As shown in \cite{wang:tcas}, we also observed that reliability variation with temperature is far more dominant than other effects. So we will use this worst-case reliability at $-45^\circ$C for our analysis going forward. The reliability of this NN-PUF without recurrence can be denoted $R_{conv}$--this notation will be useful when we talk about reliability of the RNN-PUF.

\begin{figure}
\centering
\includegraphics[clip=true,scale=0.45]{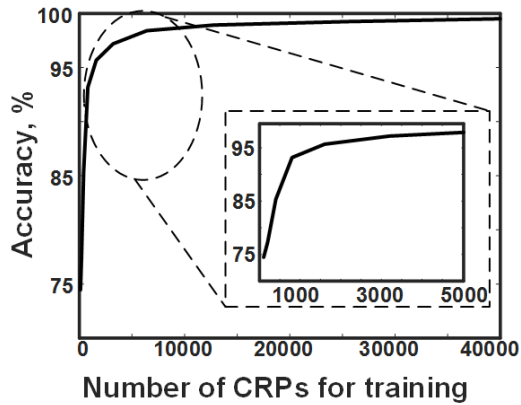}
\caption{Conventional PUF ML Accuracy - Linear SVM}
\label{convml}
\end{figure}

\subsection{Conventional NN-PUF ML Attack Accuracy}
The idea of using ML to predict responses of PUFs is well known \cite{ruhrmair:acm,lim:mit}. We attempt to use different ML algorithms to predict the response to a given challenge. To assess ML resistance of the basic structure, initially the column pair part of the challenge is kept fixed i.e. it is not randomized for different row part of the challenge. Using Linear Support Vector Machine(LSVM), we obtained the results shown in Figure \ref{convml}. Since the accuracy exceeds the reliability for $800$ CRPs, we can use this number to denote the secure space of the conventional PUF \cite{hospodar:wifs}. The results are not surprising because the conventional current-mirror-based PUF, for a fixed column pair, is similar to the APUF that has been accurately modeled before. The linear relationship between challenge and response is easily learned by the LSVM with very few training CRPs. This result shows that even analog PUFs are prone to ML attacks and there is nothing inherently secure about analog PUFs--careful design is necessary.

Another problem with conventional PUFs such as that shown in \cite{supreet:vlsi,wang:tcas} is the low Hamming Distance Test \cite{nguyen:acm} value of the output due to the comparator's bias, which biases the output toward 0 or 1, and makes it easier to predict the response. For the NN-PUF in \cite{wang:tcas}, this can be modeled as a gain factor $k_X$ multiplying the output current $I^C_{col,X}$ of the X-th column.

\begin{figure}
\centering
\includegraphics[clip=true,scale=0.4]{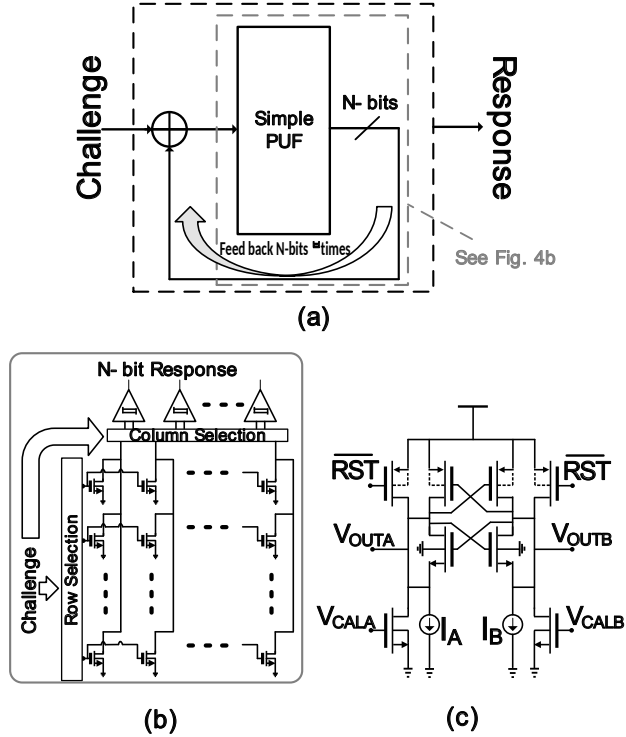}
\caption{(a) Recurrent Neural Network PUF concept (b) Circuit-level RNN-PUF core (c) High speed comparator with offset cancellation}
\label{rnnpuf}
\end{figure}

\section{Proposed RNN-PUF}
\label{sec:rnnpuf}
In order to improve over a conventional PUF's attack susceptibility, we propose the design of an RNN-PUF, whose concept is shown in Figure \ref{rnnpuf}(a). Feed-forward Neural networks such as Extreme Learning Machine have been implemented in hardware before \cite{wang:tcas} and used in dual-mode as a PUF. 
Our hypothesis was if we use the idea of Recurrent Neural Networks for a PUF, it may improve resistance to Machine Learning attacks by introducing non-linearity in the CRP space through the feedback mechanism. However, this means that one needs as many fed-back bits as there are number of challenge bits. A high number of output bits to be fed-back increases both power and area, and also reduces reliability. In order to avoid these issues, we propose another improvement wherein we XOR the feedback bits with the challenge bits.

The core of the RNN-PUF is an array of current mirrors with a shared comparator at the output, as shown in Figures \ref{rnnpuf}(b) and \ref{rnnpuf}(c). If there are $A$ rows of current mirrors and $N$ bits are fed back, then each feedback bit is XOR-ed with $A/N$ challenge bits to produce the new challenge after feedback. By sharing the comparator, we can drastically cut down power consumption compared with individual comparators for each column pair. Additionally, by using offset-cancellation we can remove the comparator bias. Another free parameter for the RNN-PUF is $\theta$, the number of times recurrence is applied before producing the final output bit. We will show a design space exploration for choosing $N$ and $\theta$ in Section \ref{sec:guide}.

\subsection{RNN-PUF ML Attack Accuracy}
To make a fair comparison, we first show MLAA reduction by using a shared comparator for the RNN-PUF as opposed to using CCO for each column pair for the conventional PUF. For these results, we randomize the column pair selection and disable recurrence, i.e., it is just a conventional PUF but without the comparator's bias. The results are shown in Table \ref{table:share} and are generated using $50,000$ CRPs using 5-fold cross-validation. There is considerable improvement by averaging out the comparator's bias, and we get a much stronger PUF against ML attacks.

\begin{figure}
\centering
\includegraphics[clip=true,scale=0.35]{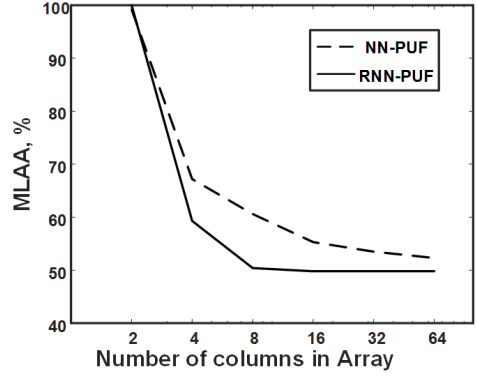}
\caption{MLAA (Boosted Trees) comparison for NN-PUF and RNN-PUF, 25,000 CRPs for each}
\label{mlaacomp}
\end{figure}

\begin{table} [b]
\centering
\caption{ML attack accuracies: conventional NN-PUF with and without comparator bias}
\begin{tabular}{|c|c|c|} \hline
 &W/ comp. bias&W/o comp. bias\\ \hline
Deep NN & 71.14\%& 52.74\%\\ \hline
Linear SVM & 51.3\% & 50.7\%\\ \hline
Bag/Boost Trees & 63.5\%& 50.9\%\\
\hline\end{tabular}
\label{table:share}
\end{table}

\begin{figure} [t]
\centering
\includegraphics[clip=true,scale=0.4]{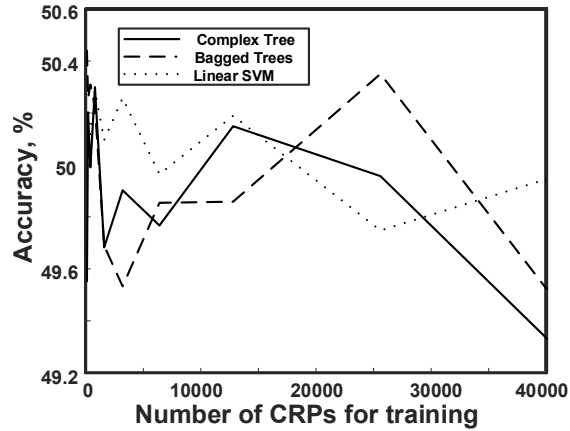}
\caption{MLAA for $128\times128$ RNN-PUF with $\theta$=1 and $N$=2 vs. number of training CRPs.}
\label{rnnacc}
\end{figure}

Now we enable recurrence for the RNN-PUF and use a shared comparator to remove bias. Although the NN-PUF with randomized column pairs is very resistant to MLAA, Figure \ref{mlaacomp} shows that we can design the RNN-PUF ($N=2$, $\theta=1$) with much fewer columns of current mirrors (and hence much smaller area) for the same MLAA compared with the NN-PUF. As with the conventional PUF ML tests, we fix the final column pair selection while intermediate cycle column pair selections are random. 
MLAA's for several algorithms are shown in Figure \ref{rnnacc}, for $50,000$ CRPs and $V_{ctrl}=0$. An accuracy of close to 50\% means that the RNN-PUF (with $N=2$, $\theta=1$) is very resilient to ML attacks and the predictability of the response is a 50-50 guess. Higher order SVMs such as Cubic SVM also show MLAA of around 50\%, so their results are not shown here. Advanced ML algorithms such as Ensemble Classifiers or Boosted/Bagged Trees are known to be more effective at breaking strong PUFs \cite{ruhrmair:acm,arun:host} compared to SVMs. However even these algorithms are unable to break the RNN-PUF, showing the effectiveness of the nonlinear structure due to recurrence. These results are all the more effective because even if a hacker has physical access to the RNN-PUF, he wouldn't be able to access the intermediate output of the RNN-PUF, without tampering with and destroying the PUF circuitry. As a result trying to model the RNN-PUF by getting the intermediate bits would be difficult even with sophisticated equipment.

It would be useful to learn how the RNN-PUF compares with other PUFs that have had non-linearity introduced into them. Two such PUFs are the XOR APUF and Feed-Forward APUF(FF-APUF), both of which have been attacked using ML techniques \cite{ruhrmair:acm,gass:ccpe,suh:dac}. In the $x$-XOR APUF, $x$ APUF outputs are XOR-ed, and one response bit is generated. This is like XOR-ing several column pair outputs of the conventional NN-PUF; hence, it is not similar to our RNN-PUF since we have both XOR-ing and cascading operations. In the FF-APUF, $x$ bits are generated at random locations within an APUF, and are then fed as part of the challenge bits to the rest of the APUF delay chain. This is comparable to feeding back output bits of the conventional NN-PUF as the challenge, without XOR-ing. Thus, it is safe to say that no attacks have been performed on APUF-like structures that use both XOR-ing and cascading, which is what the RNN-PUF is based on. A PUF construction that's closest to the RNN-PUF is the composite PUF in \cite{sahoo:host}, but no modeling attacks have been discussed.

\subsection{RNN-PUF Reliability}
\begin{figure}
\centering
\includegraphics[clip=true,scale=0.35]{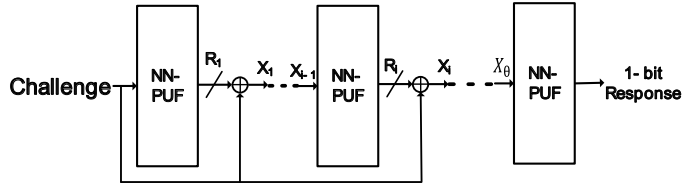}
\caption{RNN-PUF modeled as composition of NN-PUF}
\label{rnncomp}
\end{figure}

Reliability of the RNN-PUF can be decomposed into the reliability of an XOR and cascade of PUFs \cite{sahoo:host}. Figure \ref{rnncomp} shows the RNN-PUF as a composition of PUFs. This idea will help us relate the RNN-PUF reliability to a conventional PUF's reliability. We first analyze reliability mathematically for XOR operation and for cascade operation separately, and then combine them for the final analysis. XOR and cascade reliabilities are given by \cite{sahoo:host}:


\begin{equation} \label{eq:1}
R_{XOR} = R_{1}R_{2} + (1-R_{1})(1-R_{2})
\end{equation}


\begin{equation} \label{eq:2}
R_{cas} = R_{1}R_{2}
\end{equation}

where $R_1$ and $R_2$ are individual reliabilities of any process/variable. We analyze the reliabilities as a function of $N$ and $\theta$, where N is the number of intermediate bits feedback, and $\theta$ is the number of recurrences.For $R_{XOR}$, $R_1$ is the reliability of a bit feedback from the PUF, while $R_2$ is the reliability of the challenge bits, which is just 1. So $R_{XOR}$ just reduces to $R_1$. For $R_{cas}$, assuming $\theta=1$ and $N=1$, $R_1$ is the reliability of the PUF output, while $R_2$ is the reliability of the PUF behind in the chain, which is just $R_{XOR} = R_1$. So the final reliability is just $R_{cas}=R_1^2$.

Therefore, the theoretical reliability can be modeled by a recursive equation:
\begin{equation} \label{eq:3}
R_{\theta}^{N} = (R_{\theta - 1}^{N})^{N}.R_{conv}
\end{equation}

where $R_{conv}$ is the reliability of the NN-PUF, and $(R_{\theta - 1}^{N})^{N}$ is the reliability of N-bits after $\theta - 1$ cycles. This is the theoretical lower bound limit of the reliability. In practice, the reliability is observed to be higher because there is a chance that the final output is the same, even though the intermediate bits are different. Ideally, the probability that the final output is correct despite incorrect intermediate bits is $0.5$. This notion can be expressed with another equation where a correction term is added:
\begin{equation} \label{eq:4}
R_{\theta}^{N} = (R_{\theta - 1}^{N})^{N}.R_{conv} + 0.5.(1-(R_{\theta - 1}^{N})^{N}))
\end{equation}

\begin{figure}
\centering
\includegraphics[clip=true,scale=0.3]{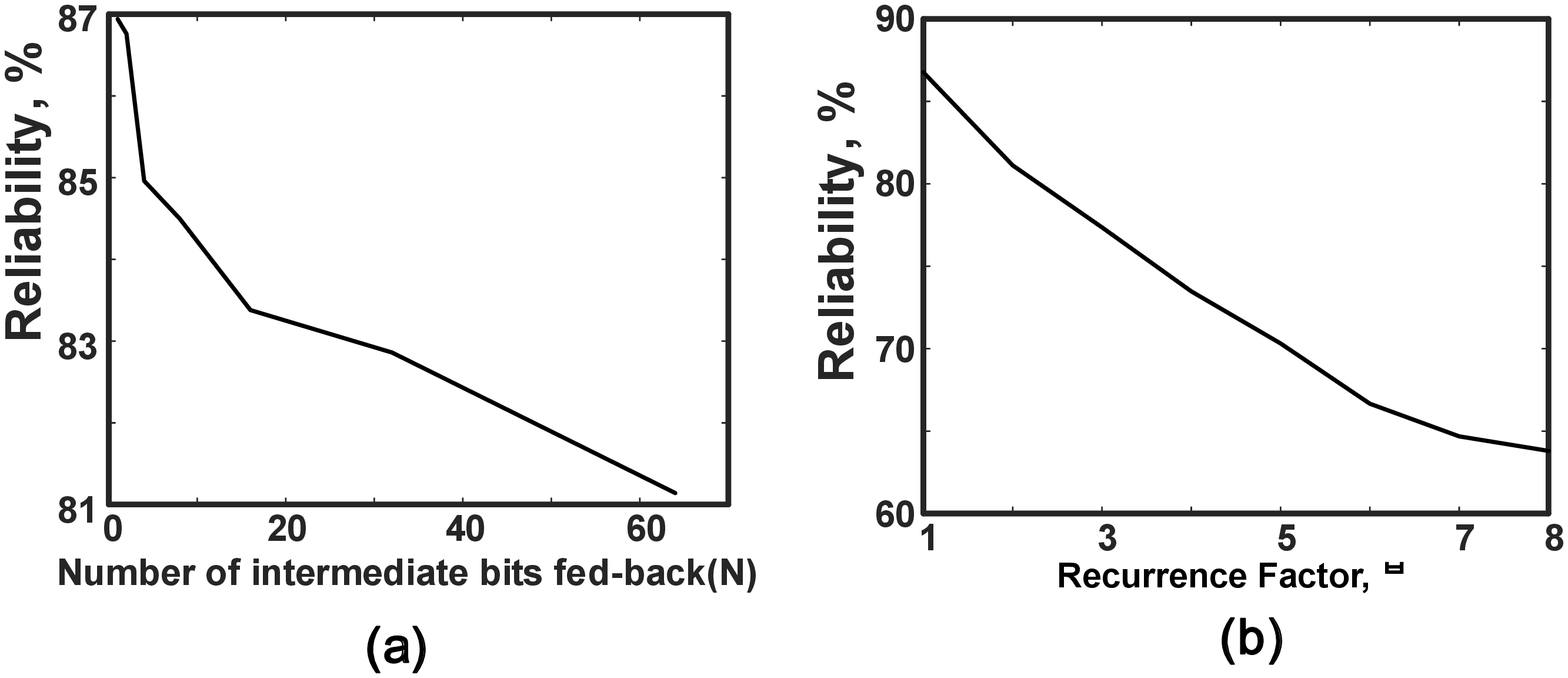}
\caption{(a) Reliability vs. $N$ for $\theta=1$. (b) Reliability vs. $\theta$ for $N=2$.}
\label{rnnrelnth}
\end{figure}

where the second factor with the multiplier of $0.5$ denotes the case when the final output is correct even though intermediate bits are incorrect. We can improve the reliability by increasing $V_{ctrl}$, but at the cost of reducing the CRP secure space. We take care of optimizing this trade-off in our Design Guide in section \ref{sec:guide}.

\begin{figure}
\centering
\includegraphics[clip=true,scale=0.48]{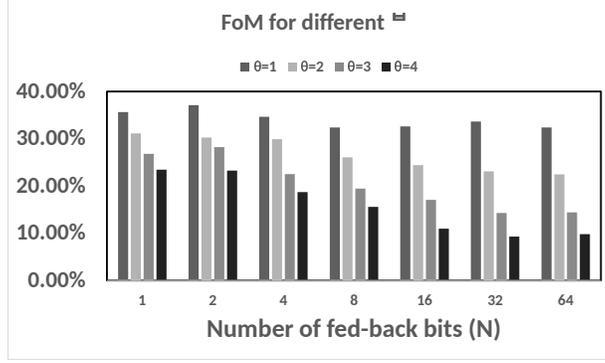}
\caption{FoM comparison of $128\times128$ RNN-PUF for different $\theta$ and $N$ values.}
\label{metric1}
\end{figure}

Figures \ref{rnnrelnth}(a) and \ref{rnnrelnth}(b) show reliability as a function of $N$ and $\theta$. When choosing $N$, we make sure that $N$ is less than half the number of columns, so that we are always choosing independent column pairs to avoid low Hamming-Distance-Test (HDT) values. We observe that the reliability drops as $N$ or $\theta$ are increased--so we should just choose low values of $N$ and $\theta$. We use these results in Section \ref{sec:guide}. Recurrence causes the reliability of the RNN-PUF to drop compared with that of the conventional PUF, however the gain in ML attack accuracy far exceeds the loss in reliability.

\section{Design Guide for RNN-PUF}
\label{sec:guide}
The RNN-PUF has $5$ parameters that we can control: $N$, $\theta$, $V_{ctrl}$, number of Rows ($A$) and Columns ($B$). The FoM used to compare the effect of parameters is the difference between reliability and MLAA:


\begin{equation} \label{eq:5}
FoM = MLAA - Reliability
\end{equation}

Figure \ref{metric1} shows the RNN-PUF's FoM for different $N$ and $\theta$. We notice that MLAA immediately drops to around 50\% for $N=1$ and $\theta$=1 for 128x128 RNN-PUF with the reliability being the highest for these parameters. Increasing $N$ or $\theta$ decreases reliability--so increasing them does not improve the FoM too much as we see in Figure \ref{metric1}. As a result it is better to go with low value of $N$, either 1 or 2, and $\theta$=1. In this way, power consumption will also stay low since only one or two column pairs are being used. We recommend using $N=2$ since there is a higher chance that the intermediate challenge is different from the original one and hence more difficult to predict.

\begin{figure} [t]
\centering
\includegraphics[clip=true,scale=0.3]{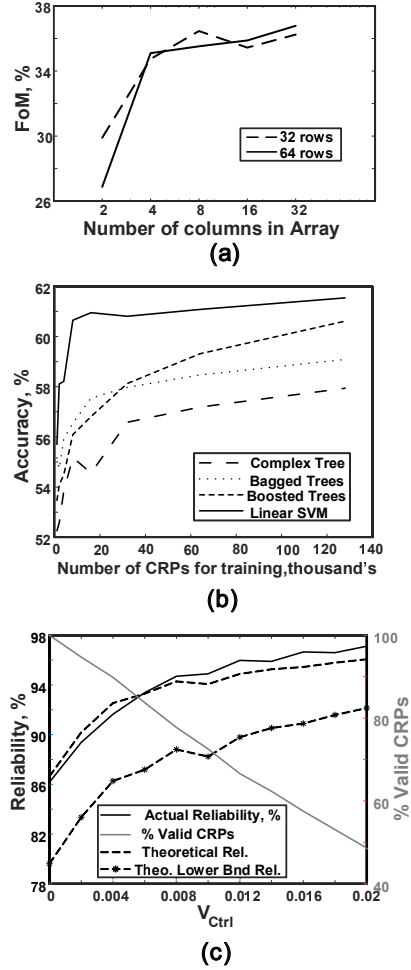} 
\caption{(a) Trade-off area vs. secure space ($\theta$=1 and $N$=2) (b) MLAA for 64x8 RNN-PUF with $\theta$=1 and $N$=2 (c) Reliability-$V_{ctrl}$ plot for 64x8 RNN-PUF with $\theta$=1 and $N$=2.}
\label{trdof}
\end{figure}

If we keep the number of rows $A$ fixed and increase number of columns $B$, we see a clear trend of decreasing MLAA in Figure \ref{trdof}(a). Since reliability stays almost the same, so the FoM increases. However increasing the number of columns increases area. Beyond $B=8$ columns, the MLAA already drops to around $50\%$, and with reliability being the same as just $2$ columns, we note that $8$ columns give sufficient performance.

Assuming $B=8$, the number of rows needs to be at least $A=32$. This is because for $A=16$, there will only be $\sim$ 262,000 CRPs which can be cracked in few days using brute force method. If we go with $A=32$, we do get a high number of CRPs, but in order to get high reliability our $V_{ctrl}$ needs to be around 0.015 which means 
we lose 25\% of the CRPs. After losing these CRPs, we have secure space of $\sim 10^{10}$ CRPs which may or may not be sufficient to withstand a brute-force attack, depending on how quickly the CRPs can be collected.

\renewcommand{\thefootnote}{\arabic{footnote}}

\begin{table*} [t]
\caption{Comparison Table}
\begin{minipage}{\textwidth}
\def\arraystretch{1}\tabcolsep 1pt
\renewcommand*{\thefootnote}{\alph{footnote}}
\footnotesize
\begin{center}
\begin{tabular}{|c|c|c|c|c|c|c|c|} \hline
 &ASPDAC'18 \cite{ma:aspdac}&FF-Arb \cite{ruhrmair:acm}&XOR-Arb \cite{ruhrmair:acm}&VLSI'17 \cite{supreet:vlsi}&VLSI'17 \cite{orshansky:vlsi}&TCAS'17 \cite{wang:tcas}&This Work\\ \hline
Array Size & 32x2 & 72x2 & 128x3 & 64x64 & (5x13)x2 & 128x128 & 64x8\\ \hline
Temp. Range, C & - & 27 to 70 & - & 0 to 80 & -20 to 80 & -45 to 90 & -45 to 90\\ \hline
Reliability, \% & - & 90.2 & - & 89 & 88 & 92.5 & 92\footnotemark\\ \hline
MLAA, \% & 80 & 95.5 & 99 & 77\footnotemark & 60 & 95\footnotemark & 61\\ \hline
FoM, \% & - & -5.3 & - & 12 & 28 & -2.5 & 31\\ \hline
Energy/bit, pJ & - & - & - & 0.097 & 11 & 3.36 & 0.16\\ \hline
Technology & FPGA & 0.18um & FPGA & 28nm & 0.13um & 0.35um & 65nm\\
\hline
\end{tabular}
\end{center}
\footnotetext[1]{\scriptsize Worst-case native reliability at 80C for fair comparison with \cite{supreet:vlsi} and \cite{orshansky:vlsi}, without any dynamic thresholding}
\footnotetext[2]{\scriptsize For 10,000/64 = 156 CRPs/column, from Fig. 9 in \cite{supreet:vlsi}}
\footnotetext[3]{\scriptsize For 200,000/64 = 3125 CRPs/col. pair from Fig. \ref{convml} in this paper}
\end{minipage}
\label{tbl:comptab}
\end{table*}

Furthermore, by having $A=32$, we do not get sufficient entropy \cite{arun:host}. Figure \ref{trdof}(a) compares the FoM for $A=32$ and $64$. We can see that the FoM does not degrade for $A=64$ due to worse reliability. As a result, we can confidently choose $64$ rows. In order to have enough CRPs, we go for 64x8 RNN-PUF with $N=2$ and $\theta=1$. Using Figure \ref{trdof}(c) we pick $V_{ctrl}=0.01$--so we lose about $28\%$ of the $191,000$ CRPs to achieve reliability of $95\%$. MLAA for the remaining $138,000$ CRPs using several algorithms is around $61\%$ at best, as seen in Figure \ref{trdof}(b), implying the FoM is still a respectable $36\%$. Figure \ref{trdof}(c) also shows the theoretical reliability plotted using Equation \ref{eq:4} and the lower bound using Equation \ref{eq:3}. We can observe that Equation \ref{eq:4} accurately models the actual reliability. In these equations, we use the fact that $R_{\theta=0}=R_{conv}$, so we get $R_{\theta=1}=R_{conv}^3+0.5.(1-R_{conv}^2)$.

\section{Power Consumption \& Comparison}
\label{sec:pwr}
We estimate power consumption of the RNN-PUF by simulating the 64x8 array of current mirrors, shared comparator and $64$ XOR gates.Technology used is UMC 65nm with 1.2V supply voltage. Since the comparator is shared, $N=2$ and $\theta=1$, the comparator burns power three times before we get the final output. All the $64$ XORs are operated in parallel; hence, latency for this operation is just that for 1 XOR gate. Total power consumed is 12.3$\mu W$, of which the array and comparator consume 11.6$\mu W$, and the XORs just 0.7$\mu W$. The clock frequency used is $250$MHz. From the time of reset to obtaining the final output bit after recurrence takes about $13$ns. As a result, we obtain an approximate Energy/bit value of $0.16$pJ/bit.

Table \ref{tbl:comptab} compares our results with recent work. Ref. \cite{orshansky:vlsi} shows good ML resistance, but the results are only for 10,000 CRPs as opposed to almost 200,000 CRPs collected for our work, for a wider temperature range. Additionally, the RNN-PUF consumes less energy and has a higher FoM of 31\% compared to 28\% at worst case temperature of 80$^\circ$C. For a fair comparison with the RNN-PUF using $8$ columns, we have deduced the MLAA's of \cite{supreet:vlsi} and \cite{wang:tcas} for the whole array, rather than just single column pair. For a multi-column/column pair PUF we can use several ML attackers in parallel, one for each column/column pair because each column/column pair is easily attacked, and therefore crack the whole PUF.

\section{Conclusions}
\label{sec:conc}
We have presented a Recurrent Neural Network strong PUF that is more resistant to ML attacks compared with conventional PUFs. Recurrence along-with XOR operation introduces non-linearity which makes it difficult even for advanced ML algorithms like Ensemble Classifiers to model the RNN-PUF. We analyzed the trade-offs and came up with a design guide to optimally design the RNN-PUF. Reducing MLAA came at a cost of reducing reliability, but the gap between reliability and attack accuracy is large and positive enough to choose the RNN-PUF over a conventional PUF. Power consumption is kept low due to sub-threshold operation of the current mirrors, making the RNN-PUF attractive for use in IoT devices. For future work, we will be implementing the RNN-PUF as a mixed-signal ASIC and will try using RNN algorithms to perform ML attack on the PUF.

\bibliographystyle{unsrt}

\end{document}